\documentclass{elsart}
\usepackage{graphicx}
\usepackage{rotating}
\usepackage{amssymb}

\newcommand{\be}{\begin{equation}}
\newcommand{\ee}{\end{equation}}

\begin{document}
\begin{frontmatter}

\title{Dynamic Peer-to-Peer Competition}
\author{L. F. Caram$^{a}$, C. F. Caiafa$^{a,b}$, A. N. Proto$^{a,c}$ and M.
Ausloos $^{d}$}
\address{
$^{a}$Lab. de Sistemas Complejos, FI-UBA.
Av. Paseo Col\'on 850, C1063ACV, Buenos Aires, Argentina.\\
\ $^{b}$ LABSP, RIKEN Brain Science Institute, Wako, Saitama 351-0198, Japan.\\
\ $^{c}$ Comisi\'on de Investigaciones Cient\'ificas PBA(CIC).
Av. Paseo Col\'on 751, C1063ACV, Buenos Aires, Argentina.\\
\ $^{d}$ GRAPES@SUPRATECS.
B5a, ULG, Sart Tilman, B-4000, Li\`ege, Belgium.
} \ead{fcaram@fi.uba.ar}

\begin{abstract}
The dynamic behavior of a multiagent system in which the agent size $s_{i}$
is variable it is studied along a Lotka-Volterra approach. The agent size
has hereby for meaning the fraction of a given market that an agent is able
to capture (market share). A Lotka-Volterra system of equations for
prey-predator problems is considered, the competition factor being related
to the difference in size between the agents in a one-on-one competition.
This mechanism introduces a natural self-organized dynamic competition among
agents. In the competition factor, a parameter $\sigma$ is introduced for
scaling the intensity of agent size similarity, which varies in each
iteration cycle. The fixed points of this system are analytically found and
their stability analyzed for small systems (with $n=5$ agents). We have
found that different scenarios are possible, from chaotic to non-chaotic
motion with cluster formation as function of the $\sigma$ parameter and
depending on the initial conditions imposed to the system. The present
contribution aim is to show how a realistic though minimalist nonlinear
dynamics model can be used to describe market competition (companies,
brokers, decision makers) among other opinion maker communities.
\end{abstract}


\end{frontmatter}%

\section{Introduction}

\begin{figure} 
\begin{center}
\includegraphics[width=14cm]{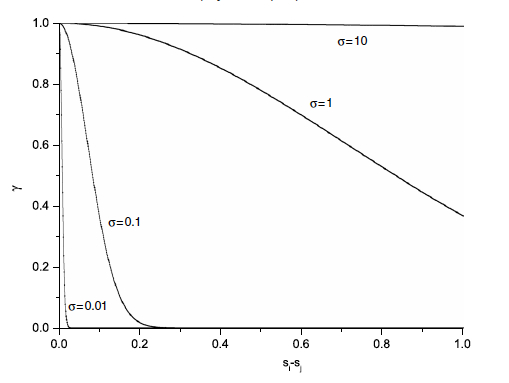}\\[0pt]
\end{center}
\caption{Interaction function $\protect\gamma $ vs. agent difference in size 
$\left( s_{i}-s_{j}\right) $, for different scaling similarity parameter ($%
\protect\sigma $) values.}
\label{figure 1}
\end{figure}

\begin{figure} 
\begin{center}
\includegraphics[
width=14cm
]{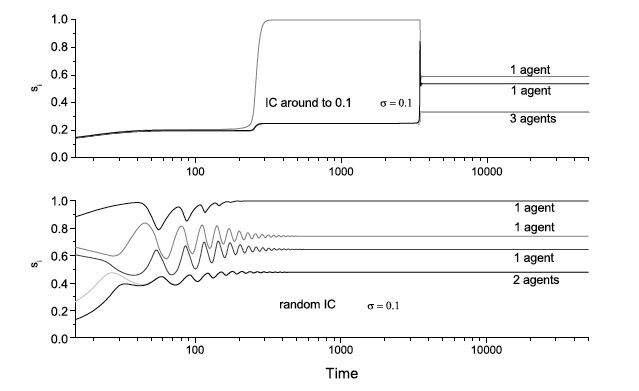}
\end{center}
\caption{Two possible simulations of the time evolution of the size of
agents for two different sets of initial conditions (IC), both for $\protect%
\sigma=0.1$, are shown. On top, it is shown when agent sizes are initially
close to each other (Table 1, case IV). On bottom, it is shown for widely
spread IC (Table 1, case VI).}
\label{figure 2}
\end{figure}

\begin{figure} 
\begin{center}
\includegraphics[
width=14cm
]{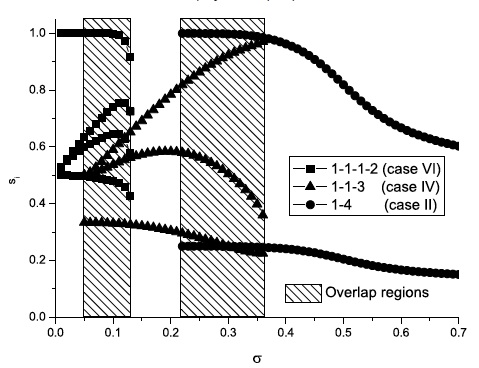}
\end{center}
\caption{Agent sizes $s_{i}$\ vs. $\protect\sigma$ values, in the $n=5$
case. In the overlap region $0.05\leq\protect\sigma\leq0.13$, cases VI and
IV coexist. Similarly in the overlap region $0.22\leq\protect\sigma\leq0.36$%
, cases IV and II coexist; see notations in Table 1.}
\label{figure 3}
\end{figure}

\begin{figure} 
\begin{center}
\includegraphics[
width=14cm
]{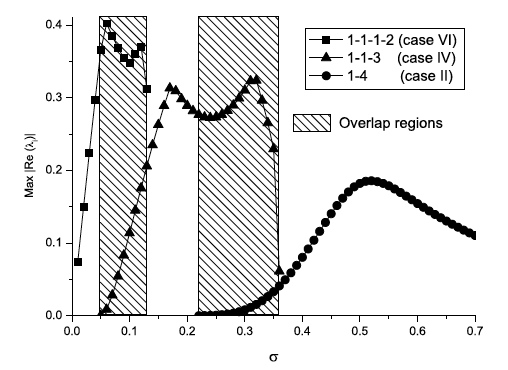}
\end{center}
\caption{Maximum absolute of eigenvalue real part of the Jacobian matrix vs. 
$\protect\sigma$ value when $n=5$. The overlap regions $0.05\leq\protect%
\sigma\leq0.13$ of eigenvalues for cases VI and IV, and $0.22\leq\protect%
\sigma\leq0.36$, corresponding to cases IV and II are shown. See notes in
Table 1 for following the case properties.}
\label{figure 4}
\end{figure}

\begin{figure} 
\begin{center}
\includegraphics[
width=14cm
]{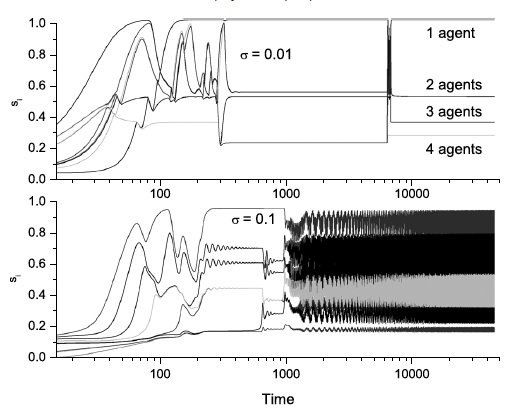}
\end{center}
\caption{Simulation of the time evolution of the size of 10 agents for two
values of $\protect\sigma$. On top, it is shown for $\protect\sigma=0.01$,
there are four clusters each containing 1, 2, 3 and 4 agents respectively.
On bottom, it is shown for $\protect\sigma=0.1$, see the change in
asymptotic behavior when $\protect\sigma$ grows, - from top to bottom.}
\label{figure 5}
\end{figure}

\begin{figure} 
\begin{center}
\includegraphics[
width=14cm
]{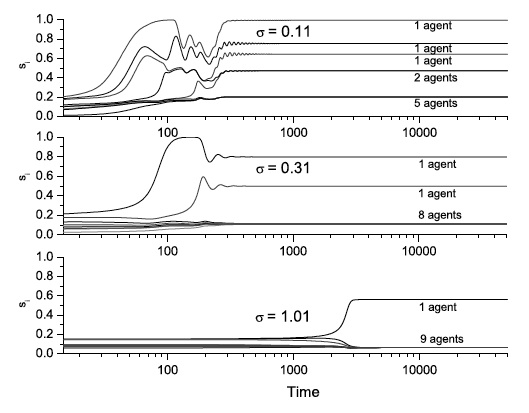}
\end{center}
\caption{Simulation of the evolution of the size of ten agents for three
different $\protect\sigma$ values, greater than in Fig.\protect\ref{figure 5}%
, i.e. $\protect\sigma= 0.11, 0.31,$ and $1.01$. Observe that the amount of
agents can vary on various levels, though keeping the total sum of agents
equal to 10; the oscillatory asymptotic behavior has disappeared}
\label{figure 6}
\end{figure}

Multiagent systems and complex networks are very active and growing areas of
research \cite{Barabasi,Amaral,Knyazeva1} with applications found in a
diverse variety of problems found in e.g. engineering systems \cite%
{ComplexDynComm}, biological systems \cite%
{Biological,Arrowsmith,Bagni,Amaral2}, neural networks \cite{NeuralNetworks}%
, socioeconomic systems \cite%
{CaiafaContrarians,Huberman,Huberman2,NetMisk,gligorMA,Redelico}.

Let it be recalled that growth, innovation pace of life in cities are scaled
using the population size \cite{Bettencourt} as urban metrics. Similarly,
biological metrics such as number, size and metabolic rate of cells are also
scaled using body sizes \cite{Savage,Moran,sawada}. In multimarket
economies, size is also relevant beside the number of competitors, entry
barriers, etc. \cite{Pilloff}. Thus many phenomena, outside well known
condensed matter physics, related to size, appear in areas that, at first
sight, do not have something in common; like sociology, biology, economy,
etc., but for which similar mathematical models can be used in fact \cite%
{Economo,Proto}.

In complex networks, individuals can behave in two opposite ways: as
cooperators or defectors as stated in the N-person prisoner dilemma \cite%
{Hauert}. Recently much attention has been given to the emergence of
"cooperation" in public goods games \cite{Santos,Glance}, and more generally
in "evolution theory". As has been used for conflict resolution according to
game theory \cite{vonNeumannandMorgenstern}, as well as reference \cite%
{MarchandSimon,Simon2} in the so-called theory of the organization. There,
they treat the conflict between groups within an organization and conflict
between organizations, these conflicts, which could be interpreted as a
situation of competition for example, are produced by: conflicting
objectives and have different perceptions of reality. So the conflict occurs
when an agent or agent group experiences a decision problem. Negotiation and
politics are the procedures used by the organization to manage these
conflicts and lead to tensions in status and power systems. If those who are
formally more powerful prevail, they give rise to a stronger perception of
differences in status and power, if not sustained its position is weakened.
To address the conflict between organizations has used the theory of
negotiation, albeit with a gap that empirical and negotiation situations are
so complex that it is not possible to develop a theory general. Within this
framework we can also quote \cite{Porter,Porter2}. Which defines five forces
that determine the competitive strategy of an organization, within them the
bargaining power and market threats result in the rivalry of existing
competitors. Finally one of the latest Nobel laureate in economics, \cite%
{Coase,CoaseWilliamson,Williamson} referred to the formation of firms,
companies, rather than individual goods and services marketed by itself,
this is given by the structure of transaction costs.

In the present paper, we focus on a competitive scenario based on a set of
differential equations as it will be explained in Section 2. In short, we
start from a Lotka-Volterra set of equations for describing prey-predator 
\cite{pekalskirevPP} situations \cite{Arrowsmith,Lotka}, i.e. a system of $n$
agents competing for some common resource. An original constraint is imposed
on the agent dynamics, i.e. an increase in size is favored in order to
obtain resources, while an agent decreasing size implies the loss or lack of
getting such resources. By size we mean something like the fraction of a
given market that an agent has, i.e. its market share in economy. Other
intuitive notions of size are easily imagined in many other systems.

Let us mention that in \cite{Huberman}, a different generalization of the
Lotka-Volterra model \cite{Lotka} to the $n$ agent (competitor) case was
used for describing the information electronic web's dynamics. Two behavior
states were found depending on a global interaction parameter: on one hand,
a stationary state exists in which only one agent gets all the available
resources ("winner takes all") \cite{Huberman,Huberman2}, while the rest of
agents gets nothing; on the other hand, there is another stationary state
where the resource is fairly shared among all agents ("sharing the market").
In several works \cite{CaiafaContrarians,Hauert,Szabo,Szabo2,Caram}, spatial
configurations of agents were also considered allowing them to interact only
with a limited set of neighbors, thereby resulting in spatially distributed
patterns as stationary states of the system \cite{CaiafaContrarians}. Let us
also finally cite \cite{Yanhui} in which the interaction coefficient of the
model of \cite{Huberman} has a linear form, giving rise to the effect
"riches get richer".

Practically we introduce an interaction function (IF) in the Lotka-Volterra
model which depends on an agent size in order to let the agents compete
among themselves in a self-adjusting way. This contrasts with assuming a
fixed interaction coefficient like in the previously cited references. That
means that the competition process and outcome are now intertwined variables
since they are related to changes in the size of agents. A parameter $\sigma$
that governs the spread in competition strength, i.e. for scaling the
difference between the size of agents, is also introduced. The IF
modification will be shown to lead to a dynamics of the system characterized
by a time dependent competition. As $\sigma$ is kept fix along each
simulation, we are able to see how different agent groups compete among
themselves. In short, this means that the IF is varying with time,
accordingly with the size of agents is changing, implies an intrinsically
novel dynamics of competition.

This paper is organized as follows: in Section 2, the mathematical model is
outlined; in Section 3, the system fixed points are theoretically analyzed
and an illustrative example for a system with a small number (five) of
agents is provided; in Section 4, simulation results are presented for the
case of a larger number (ten) of agents, in order to establish the
qualitative validity of the quantitative findings obtained for the small
size systems. This allows us in Section 5, to outline the novelty of this
work and emphasize some conclusion.

\section{Peer-to-peer competition model}

Let us consider a system with $n$ agents competing for some common resource,
as in a generalized predator-prey model or Lotka-Volterra model studied in 
\cite{Huberman}. On this model, a competition parameter, $\gamma$, is
introduced, which allows the agents to compete among themselves in a
self-adjusting way. The system is governed by the following set of $n$
differential equations:%
\begin{equation}
\dot{s}_{i}=\alpha_{i}s_{i}\left( \beta_{i}-s_{i}\right) -{\displaystyle%
\sum\limits_{i\neq j}} \gamma\left( s_{i},s_{j}\right) s_{i}s_{j} 
\mbox{ \ \
\ \ \ \ for \ \ \ \ } i=1,..,n  \label{Model}
\end{equation}
where $s_{i}$ is the size of agent $i$ in the range $0<s_{i}\leq1 $; $\dot{s}%
_{i}$ is its time derivative; $\alpha_{i}$ is the growth rate of agent $i$
if no interaction is present; $\beta_{i}$\ is the maximum capacity of agent $%
i$ and $\gamma\left(s_{i},s_{j}\right) $ is the IF hereby defined by:%
\begin{equation}
\gamma\left( s_{i},s_{j}\right) =\exp\left[ -\left( \frac{s_{i}-s_{j}}{\sigma%
}\right) ^{2}\right]  \label{gamma}
\end{equation}
in which $\sigma$ is a global positive parameter which controls/scales the $%
s_{i}$ degree of similarity in the competition.

Notice that the analytical form in Eq.(\ref{gamma}), resembling the Gauss
function used in statistics, has been chosen because it allows both
analytical and numerical approaches; it has many attractive mathematical
properties, i.e., it is a continuously differentiable function allowing us
to make a proper theoretical analysis of the system dynamics. In fact, one
could simply require any positive and even function of the absolute
difference of agent sizes $\Delta=\left\vert s_{i}-s_{j}\right\vert $ with
the property that its maximum is located at $\Delta=0$ and it is a
decreasing function for positive $\Delta$. We suggest that another case
could be the Kac potential \cite{Kac}. Other forms can be imagined but to
compare them is not the primary purpose of this paper.

From Eq.(\ref{gamma}), we see that the interaction is symmetric and always
positive representing a competitive and fair scenario; the competition
factor is maximum and equal to one when $s_{i}=s_{j}$. Additionally, we
observe that, as the absolute difference of agent sizes $\left\vert
s_{i}-s_{j}\right\vert $ becomes large the competition factor tends to be
small. Plots of the IF $\gamma\left(s_{i},s_{j}\right) $ versus the absolute
difference of two agent sizes for different values of $\sigma$ are shown in
Fig.\ref{figure 1} in order to suggest quantitatively reasonable values.

As we let the interaction coefficient depends on the difference in agent
sizes, a peer-to-peer competition is imposed, in the sense that agents with
a similar size compete more aggressively than agents with different sizes.
The name of peer-to-peer competition model makes sense in this context, as
only agents with similar sizes are able to compete reciprocally.

We emphasize that this kind of competition is a phenomenon that can be
observed in many socioeconomic systems, i.e. the competition is strong among
big companies while the competition is weak between a big company and a
medium or small one. Similarly for banks, hedge funds, pension funds,
countries, etc. Even in poker game, it is the money the player has at
his/her disposal which allows him/her to compete and influence the game.

As it will be shown below, different stationary states emerge as a function
of $\sigma$, which are interestingly characterized by agent grouping.
Self-organized clustering as well as different possible scenarios appear,
ranging from chaotic to non-chaotic motion, depending not only on the value
of the size similarity measure but also on the agent size initial condition%
\footnote{%
By initial condition we mean the set of initial agent size values, at
simulation time equal to zero.}.

It should be further noticed that, as $\sigma$ tends to be larger ($%
\sigma\rightarrow\infty$) the similarity of competition tends to be static,
i.e. $\gamma\left(s_{i},s_{j}\right) \rightarrow1$ constant, approaching to
the particular case of "winner takes the maximum, or almost all".

In this paper we consider that all agents have the same dynamics properties
restricting our study to the case where $\alpha_{i}=1$ and $\beta_{i}=1$ for
which the equations are:%
\begin{equation}
\dot{s}_{i}=s_{i}\left( 1-s_{i}\right) -{\displaystyle\sum\limits_{i\neq j}}
\gamma\left( s_{i},s_{j}\right) s_{i}s_{j} \mbox{ \ \ \ \ \ \ for \ \ \ \ }
i=1,..,n  \label{final model}
\end{equation}

\section{Fixed points analysis}

First, the existence of fixed points and a study of their stability are
presented. By definition, a fixed point is a point in the phase space where
all the time derivatives are zero, i.e.,%
\begin{equation}
\dot{s}_{i}=0 \mbox{ \ \ \ \ \ \ for \ \ \ \ } i=1,..,n.  \label{condicionPF}
\end{equation}

\subsection{Trivial fixed points for $n$ arbitrary agents}

From Eq.(\ref{condicionPF}), we detect at least three trivial fixed points
which are:

(I) $s_{i}=0$ \ \ \ \ \ \ for $i=1,.. , n$ (all agents with zero size);

(II) $s_{i}=1$ and $s_{j}=0$ for every $j\neq i$;

(III) $s_{i}=b$ \ \ \ \ \ \ for $i=1,..,n$ (all agents with the same size $b$%
)

In the latter case (III), we can directly calculate the corresponding
constant $b$ by using Eq.(\ref{final model}) as follows:%

$
0=b\left( 1-b\right) -(n-1)b^{2}=1-bn;
$
whence we necessarily have $b=\frac{1}{n}$.

As usual, for the analysis of a system fixed points stability, one needs to
look at the eigenvalues of the Jacobian matrix $J$ evaluated at the
corresponding fixed points. It is shown that the elements of the Jacobian
matrix for the system are:%
\begin{equation}
\left[ J\right] _{\left( i,k\right) }=\frac{\partial\dot{s}_{i}}{\partial
s_{k}}=\left\{ 
\begin{array}{ll}
1-2s_{i}-{\displaystyle\sum\limits_{i\neq j}} s_{j}\gamma\left(
s_{i},s_{j}\right) \left[ 1-\frac{2}{\sigma^{2}}s_{i}\left(
s_{i}-s_{j}\right) \right] & \mbox{\ \ for \  } k=i \\ 
-s_{i}\gamma\left( s_{i},s_{k}\right) \left[ 1+\frac{2}{\sigma^{2}}%
s_{k}\left( s_{i}-s_{k}\right) \right] & \mbox{\ \ for \  } k\neq i.%
\end{array}
\right.  \label{Jacobian}
\end{equation}

If we evaluate the Jacobian matrix at the fixed points (I), from Eq.(\ref%
{Jacobian}) we obtain the identity matrix; all its eigenvalues are equal to
one $\left( \lambda_{i}=1\right) $ and therefore it is an unstable fixed
point.

Evaluating Eq.(\ref{Jacobian}) at the second type of fixed point (II) for
the case with $s_{1}=1$ and $s_{2}=s_{3}=..=s_{n}=0$, we obtain that the
Jacobian matrix is:%

$
J=\left[ 
\begin{array}{ccccc}
-1 & -a & -a & .. & -a \\ 
0 & 1-a & 0 & .. & 0 \\ 
0 & 0 & 1-a & .. & 0 \\ 
: & : & : & .. & : \\ 
0 & 0 & 0 & .. & 1-a%
\end{array}
\right]
$

with 
$
a=\exp\left( -\sigma^{-2}\right).
$
It can be shown that the eigenvalues of $J$ in this case are:%
\begin{equation}
\begin{array}{l}
\lambda_{1}=-1 \\ 
\lambda_{2,3,..,n}=1-a=1-\exp\left( -\sigma^{-2}\right)%
\end{array}
\label{autovaloresII}
 \end{equation}

From Eq.(\ref{autovaloresII}) we can deduce that this fixed point is not
stable since it has $n-1$ positive eigenvalues; this is neither dependent on
the number of agents nor on the value of the parameter $\sigma$.

Next, we analyze the stability of the third type of fixed point (III). By
following a procedure similar to the previous one, we obtain the following
Jacobian matrix:%

$
J=\frac{1}{n}\left[ 
\begin{array}{ccccc}
-1 & -1 & -1 & .. & -1 \\ 
-1 & -1 & -1 & .. & -1 \\ 
-1 & -1 & -1 & .. & -1 \\ 
: & : & : & .. & : \\ 
-1 & -1 & -1 & .. & -1%
\end{array}%
\right]
$

whose eigenvalues are:%
\begin{equation}
\begin{array}{l}
\lambda _{1}=-1 \\ 
\lambda _{2,..n}=0%
\end{array}
\label{autovaloresIII}
\end{equation}%
which reveals that it is not a stable fixed point.\footnote{
If one wants to be more rigorous, would need to calculate the second order derivatives since the zero eigenvalues do not wholly determine the stability.
Our numerical simulations do not show this fixed point to be a stable one.}

Additionally to the fixed points of type (I), (II) and (III) there are many
other points that verify the conditions in Eq.(\ref{condicionPF}). These
points are found by seeking the roots of a set of $n$ non-linear equations.

\subsection{Trivial fixed points for a small finite number of agents}

In order to illustrate the analysis we restrict ourselves to examine the
case of $n=5$ agents. Considering the degeneracy of several solutions, seven
possible states appear in which a different combination in the number of
agents can be involved. In this sense it can be said that such final states
can be called "size levels" or in short "levels". Moreover since several
agents can be on the same level, it can be said that "clusters" of agents
are obtained. Table 1 enumerates these cases and the type of stability; the
latter has been determined by numerically evaluating the Jacobian matrix Eq.(%
\ref{Jacobian}) at the fixed point through a Newton-Raphson (NR) algorithm 
\cite{Myller}.

Let us examine the seven cases obtained through the numerical solution of
Eq.(\ref{Jacobian}). They are summarized in Table 1.

\begin{itemize}
\item Case I: only one state final size, level, is found to be occupied by
the five agents (5).

\item Case II and III two levels appear; with two distributions: called
(1-4) and (2-3) i.e. made either of 1 agent and a group of 4 agents or a
group of 2 agents and a group of 3 agents, respectively.

\item Likewise Cases IV and V are two distributions with three levels
(1-1-3) and (1-2-2) respectively.

\item Case VI has only one distribution with four levels, (1-1-1-2).

\item Finally in Case VII again only one distribution is found, now with
five levels, (1-1-1-1-1).
\end{itemize}

Notice that from Eq.(\ref{Jacobian}) the upper levels are always $less$
occupied than the lower one. This fact reflects what seems to happen in real
life, as "big players" are less numerous than small ones. Observe also that
the number of levels and the fixed point stability depend on $\sigma$.


It is also found that the result depends on the initial conditions. Indeed,
observe in Fig.\ref{figure 2}, the time evolution of the $s_{i}$ set
illustrated when $\sigma =0.1$, $n=5$ and for two different sets of initial
conditions. As it can be seen, for these cases, after 1000 iteration (time)
steps different states ("levels") $s_{i}$ values are asymptotically reached.
For arbitrary, random, IC we obtain the solution corresponding to case VI,
i.e. (1-1-1-2). However when the size initial conditions are imposed to be
taken from a narrow range, the resulting solution corresponds to the case
IV, i.e. (1-1-3). Since, as in Fig.\ref{figure 2}, two different stable
solutions can exist for the $same$ $\sigma $ value, but for two different
sets of initial conditions (IC), this sort of local degeneracy must be
stressed: we call it an overlap region. We emphasize that according to the
initial conditions the population of the final states can be markedly
different. This observation is reinforced when we analyze the agent size $%
s_{i}$ $vs.$ $\sigma $, here below.

\begin{table}
\begin{center} \begin{tabular}{|c|c|c|c|}
\hline Cases & Size at the stationary state  & Type & Stability\\
 \hline Case  1&All agents with the same size & one level&\\   &  & 5 & Non stable  \\
 \hline Case II&One  agent with one size  &two levels & \\ 
 &and the rest of agents with another size & & \\ &(trivial fixed point) & 1 - 4 & Stable depending on $\sigma$\\
 \hline Case III&Two agents with one size &two levels & \\ &and the rest of agents with another size& 2-3 & Non stable \\
 \hline Case IV   &  Two agents with one size&three levels & \\ & and three agents with the same size & 1-1-3& Stable depending on $\sigma$\\

\hline Case V  & One agent with one size, &three levels & \\ &two agents with another size & & \\ &and two agents with
another size&1-2-2 &Non stable \\
 \hline Case VI &Two agents with one size &four levels& \\ &and the rest of agents with different sizes & 1-1-1-2 & Stable depending on $\sigma$\\
  \hline  Case VII&All agents with different sizes&five levels&\\  &  & 1-1-1-1-1 & Non stable \\  \hline
\end{tabular}  
\caption{The seven cases emerging from the numerical solution of Eq. (5) by the Newton-Raphson algorithm are summarized. Cases are ordered according with the
number of final levels and their characteristic stability are enumerated in the last column. Further explanations are in the text.}
\end{center}
\end{table}

In Fig.\ref{figure 3}, the corresponding agent sizes are shown at the stable
fixed points as a function of the parameter $\sigma $. In fixed point
searching, we have used the NR method, departing from 100000 random initial
conditions looking for the stable fixed points. The numerical evaluation of
the Jacobian matrix at different $\sigma $ values has been done through the
NR procedure. Any solution that is not a stable fixed point was discarded.
So all the stable fixed point solutions are considered to be found. It can
be observed that there are $two$ overlap regions in $\sigma $ space where
more than one fixed point exists. In the overlap region $0.05\leq \sigma
\leq 0.13$, cases VI and IV coexist. In the overlap region $0.22\leq \sigma
\leq 0.36$, cases IV and II coexist. In simple words, agents with different
asymptotic in time $s_{i}$ value, or final level, can be simultaneously
found, for a given $\sigma $, thus for a given IF.

Given that absolute eigenvalues determine the strength of the attractor, we
can search for the value of the maximum absolute of eigenvalue real part of
the Jacobian matrix for each fixed point; this result is shown in Fig.\ref%
{figure 4}. A similar feature, i.e. overlap regions, as in Fig.\ref{figure 3}
is observed as a function of $\sigma$, depending on the IC. Thus it is found
that the dynamics of the system is dominated by the fixed point of the
maximum absolute of eigenvalue real part; the precise dynamics \emph{also}
depends on the initial conditions.

Recall that Cases I, III, V and VII are not stable. However Cases II, IV and
VI show an interesting type of degenerate stability (two possible solutions
coexist within specific ranges of $\sigma$) as reported in the previous
paragraph. While unique solutions are found in: (i) case VI for $\sigma
\leq0.05$, (ii) case IV for $0.13\leq\sigma\leq0.22$ and (iii) case II for $%
0.36\leq\sigma$. Many numerical investigations lead us to conclude that

\begin{itemize}
\item for random initial conditions, i.e. for a widely spread range of
initial sizes, the solution corresponding to the maximum absolute of
eigenvalue real part appears.

\item when the initial conditions are taken from a non uniform, i.e. narrow,
distribution, the solution corresponding to the other eigenvalue emerges.
\end{itemize}

\section{Simulation results for "large" systems}

In this section we present some simulation results for a system with $n=10$
agents, all of them varying between $0$ and $1$ in size. We have made
simulations where the $\sigma_{i}$ parameter has been varied in a
logarithmic way, from $0.01$ up to $1$. In summary, one can observe
different dynamics, ranging from the trivial result with almost no
competition, when $\sigma$ is small in relation to the difference of agent
sizes, passing through oscillatory regimes, cluster formation regimes, up to
the state where only one agent is the biggest and the rest of them remain at
a very low level state, when $\sigma$ is large.

To solve our differential equations system, we have used a typical algorithm
for numerical integration \cite{Myller}. The initial size of each agent in
each simulation was randomly chosen from a uniform distribution.

The system evolution for ten agents for $\sigma=0.01$ and $\sigma=0.10$ is
shown in Fig.\ref{figure 5}. When $\sigma$ grows, a change in the system
behavior occurs. For $\sigma=0.01$, four size levels having 1.0, 0.5, 0.33
and 0.25 values are obtained populated by 1, 2, 3 and 4 agents respectively.
It is found again that small size agents form more populated clusters. The
clusters are formed after a transitory regime. A similar highly complex
behavior can be seen, up to around 500 iterations for $\sigma=0.10$. For
higher time iterations some erratic, oscillations appear. No asymptotically
strictly stable in time solution exists. For very large time one can
consider that a non stable stationary state is reached; it can be
conjectured that the corresponding fixed points are non stable either. The
evolution goes toward a band clustering rather chaotic situation.

However if $\sigma$ is further increased above $0.10$ (Fig.\ref{figure 6}),
the system asymptotically reaches a stable stationary state, so that a sort
of condensation of agents occurs and a cluster type situation appears, with
a densely populated level at the lowest size. For $\sigma=0.11$, it can be
observed that there are 5 clusters (1-1-1-2-5); for $\sigma=0.31$ there are
3 clusters (1-1-8) and finally for $\sigma=1.01$ only two clusters are
present (1-9). It can be seen that agents abandon the oscillatory behavior,
seen at intermediate $\sigma$ values, in order to form different groups or
clusters with strong internal competition; this case corresponds e.g.
already to the 1-1-1-2-5 set, and is huge in the last case ($\sigma=1.01$)
when the lowest size cluster is made of 9 agents ; recall that the
competition factor is maximum and equal to one when $s_{i}=s_{j}$ (Fig.\ref%
{figure 2}).

It can be also observed that further increasing $\sigma$ induces a
systematic decrease in the number of levels, - the lowest level being always
the more crowded. Our numerical simulation has shown that the situation is
found for two levels only, i.e. $\sigma=1.01$, it emerges when $\sigma$ is
around $\simeq$ 0.5. This persists for higher $\sigma$ values. This behavior
in which one agent obtains the highest level while the rest of agents
reaches the lowest common level illustrates that some sort of monopoly has
been configured as the final state of the competition.

\section{Conclusions}

The main input contribution of this paper consists in the introduction of a
self-organized competition scheme related to agent sizes. The agent size
acts into our context, through the IF, as a limitation of the number of
possibilities for agent interaction or as a constraint. If the difference in
sizes is out of the $\sigma$ scale they are not able to interact anymore. So
they are "constrained" in some sense. This constraint allows that only
"players" of similar sizes are able to or truly compete with each other. A
quite simple example is given by a poker game when, at the end of the game,
only those with the same capacity to bet can be continuing the game. Other
cases occur in many finite size markets, like pharmaceutical companies
through drugstores, fuel distribution through gasoline stations, clothing
and specialty goods in supermarket chains.

In the present analysis of the clustering of agents no use is made of
techniques like those used physics literature on synchronization \cite%
{Ermentrout,Strogatz,Liu,Zheng1,Hu,Zheng2}, which have been around for a
couple of decades, because we do not use chaos synchronization in our
contribution. Such extensions should nevertheless be suggested. Our model
leads to describe cases when a strong self-organized feedback scheme
essentially occurs, i.e. the number of interacting agents being a somewhat
secondary aspect. As we are dealing with a nonlinear system, the initial
conditions impose severe constraints leading to specific dynamics and a
variety of final stationary states. Indeed several behaviors emerge, going
from one extreme in which a few agents compete with each other, passing
through oscillations with clustering, up to the case of a "winner takes the
top" state, and all others drop out. This latter state reflects the fact
that the IF tends to a constant value $=1$, as the size scale parameter $%
\sigma$ increases. It can be said in simple words that this corresponds to
when "the competition is at its maximum".

We have stressed the case of a small number $n$ of agents which seems the
most reasonable practical case \cite{manna,garciamolina}. The stability
analysis for five agents as presented exemplifies the different dynamics and
final state possibilities. Here it was demonstrated that there exists a
critical $\sigma$ of the system, $0.05\leq\sigma_{1}\leq0.13$ and $%
0.22\leq\sigma_{2}\leq0.36$, where the solution is no longer unique. It can
be extrapolated the present findings to other finite number $n$ of agents
cases.

As shown in several figures, a clustering phenomenon can be also obtained in
which the competition is in fact between clusters of agents. In market
language, this situation represents the natural segmentation into big,
medium, and small players. We have shown that the segmentation can be
extreme, even of the binary type. From a socioeconomic point of view, this
means that a monopolistic situation is sometimes likely.

We emphasize the relevance of initial conditions, but agree that in general
they are hard to define in socioeconomic systems.

Notice also that even when clusters (in size) appear there still exists
competition among the agents $inside$ the cluster; this competition is
stronger at the lowest size levels, which are the most densely populated.
This fact well reflects the complexity of markets: not only agents with
equal sizes are in competition with those with bigger or smaller sizes, but
also they are still in strong competition with each other at their own
level. In some sense, the only final state will be, sooner or later, only
one agent at the top, because of the attrition of the small ones.

Finally and in addition we would like to re-emphasize that the competition
scheme presented here Eq.(\ref{gamma}) gives rise to a very complex
dynamics, although based on a simple idea, and on having only one quantity $%
\sigma$, as the regulation parameter of the different behaviors. This can be
put in line with the observed complexity of evolution of true systems,
though is clearly a reductionist view of the world \cite{Knyazeva2}.

\textbf{Acknowledgements}: Thanks to a CONICET - CGRI-FNRS bilateral
agreement which has allowed interaction between the authors.


\begin{thebibliography}{99}
\bibitem{Barabasi} R. Albert, A.L. Barab\'{a}si, 
Rev. Mod. Phys. 74 (2002) 47-97. 

\bibitem{Amaral} L.A.N. Amaral, J.M. Ottino, 
Eur. Phys. J. B 38 (2004) 147-162. 

\bibitem{Knyazeva1} H.N. Knyazeva, S.P. Kurdyumov, Evolution and
Self-organization Laws of Complex Systems, Moscow, Nauka Publisher in
Russian 1994 p~236. 

\bibitem{ComplexDynComm} L. Kocarev, G. Vattay, Complex Dynamics in
Communication Networks, Springer Verlag 2005.

\bibitem{Biological} S. Sitharama Iyengar, Computer Modeling and Simulations
of Complex Biological Systems, 2nd edn. CRC, 1997.

\bibitem{Arrowsmith} D. Arrowsmith, C.M. Place, Dynamical Systems:
Differential Equations, Maps and Chaotic Behaviour, Chapman and Hall, 1992.

\bibitem{Bagni} R. Bagni, R. Berchi, P. Cariello, 
J. Artif. Soc. Social Simul. 5 (2002) 3. $%
http://jasss.soc.surrey.ac.uk/5/3/5.html$

\bibitem{Amaral2} F. Liljeros, C.R. Edling, L. A. N. Amaral, 
Microbes Infect. 5 (2003) 189-196. 

\bibitem{NeuralNetworks} Zhang Yi, K.K. Tan, 
IEEE Trans. Circuits  Syst. 52 (2005) 2482-2489. 

\bibitem{CaiafaContrarians} C.F. Caiafa, A.N. Proto, 
Intern. J. Mod. Phys. C 17 (2006) 385-394. 

\bibitem{Huberman} S.M. Maurer, B.A. Huberman, 
J. Econom. Dynam. Control 27 (2003) 2195-2206. 

\bibitem{Huberman2} L.A. Adamic, B.A. Huberman, 
Quart. J. Electron. Commerce 1 (2000) 5-12. 

\bibitem{NetMisk} J. Miskiewicz, M. Ausloos, 
in Proc. of the 3rd Nikkei Econophysics Symposium: Practical Fruits of
Econophysics, Tokyo, ed H. Takayasu, Springer 2006, p. 312. 

\bibitem{gligorMA} M. Gligor, M. Ausloos, 
Eur. Phys. J. B 57 (2007) 139-146. 

\bibitem{Redelico} F.O. Redelico, A.N. Proto, M. Ausloos, Physica A 388
(2009) 3527-3535.

\bibitem{Bettencourt} L.M.A. Bettencourt, J. Lobo, D. Helbing, C. Kuhnert,
G.B. West, 
Proc. Natl. Acad. Sci. 104 (2007) 7301-7306. 

\bibitem{Savage} V.M. Savage, A.P. Allen, J.H. Brown, J.F. Gillooly, A.B.
Herman, W.H. Woodruff, G.B. West, 
Proc. Natl. Acad. Sci. 104 (2007) 4718-4723. 

\bibitem{Moran} A.L. Moran, J.D. Allen, 
Biol. Bull. 212 (2007) 143-150. 

\bibitem{sawada} Y. Sawada, 
Physica A 204 (1994) 543-554. 

\bibitem{Pilloff} S.J. Pilloff, 
Rev. Industr. Organ. 14 (1999) 163-182. 

\bibitem{Economo} E.P. Economo, A.J. Kerkhoff, B.J. Enquist, 
Ecol.Lett. 8 (2005) 353-360.

\bibitem{Proto} N. Olivera, A.N. Proto, F. Redelico, 
RC332008 7 th International Conference on Social Science Methodology, Campus
di Monte Sant'Angelo, Naples, Italy (2008). 

\bibitem{Hauert} C. Hauert, S. De Monte, J. Hofbauer, K. Sigmund, 
Science 296 (2002) 1129-1132. 

\bibitem{Santos} F.C. Santos, M.D. Santos, J.M. Pacheco, 
Nature 454 (2008) 213-216. 

\bibitem{Glance} N.S. Glance, B.A. Huberman, 
Sci Am. 270 (1994) 76-81. 

\bibitem{vonNeumannandMorgenstern} J. von Neumann, O. Morgenstern, Theory of
Games and Economics Behavior, Princenton, 1944.

\bibitem{MarchandSimon} J.G. March, H.A. Simon, Organizations, Wiley, New York,
 1958.

\bibitem{Simon2} H.A. Simon, Administrative Behavior, Free Press, New York, 1947.

\bibitem{Porter} M. Porter, Competitive Strategy, Free Press, New York 1980.

\bibitem{Porter2} M. Porter, Competitive Advantage, Free Press, New York
1985.

\bibitem{Coase} R.H. Coase, The firm, the market, and the law Chicago:
University of Chicago Press,1988.

\bibitem{CoaseWilliamson} O.E. Williamson, S.G. Winter, R.H. Coase, The
Nature of the firm: Origins, evolution, and development New York: Oxford
University Press, 1991.

\bibitem{Williamson} O.E. Williamson, The Economic Institutions of
Capitalism. Free Press, New York, 1985.

\bibitem{pekalskirevPP} A. Pekalski, 
Comp. Sci. Eng. 6 (2004) 62-66. 

\bibitem{Lotka} A.J. Lotka, Elements of Physical Biology, Williams\&Wilkins
Co., Baltimore, 1925, p.460.

\bibitem{Szabo} G. Szabo, C. Hauert, 
Phys. Rev. Lett. 89 (2002) 118101.

\bibitem{Szabo2} G. Szabo, C. Hauert, 
Phys. Rev. E 66 (2002) 062903.

\bibitem{Caram} L.F. Caram, C.F. Caiafa, A.N. Proto, 
Intern. J. Mod. Phys. C 17 (2006) 435-445. 

\bibitem{Yanhui} Li Yanhui, Zhu Siming, 
Appl. Math. Modelling 31 (2007) 912-919. 

\bibitem{Kac} M. Kac, 
Phys. Fluids 2 (1959) 8-12. 

\bibitem{Myller} W.H. Press, B.P. Flannery, S.A Teukolsky, W.T. Vetterling,
Numerical Recipes in FORTRAN: The Art of Scientific Computing, 2nd edn,
Cambridge: Cambridge University Press, 1992,  p.364.

\bibitem{Ermentrout} G.B. Ermentrout, N. Kopell, SIAM J. Math. Anal. 15
(1984)  215-.

\bibitem{Strogatz} S.H. Strogatz, R.E. Mirollo, Physica D 31  (1988) 143-.

\bibitem{Liu} Z. Liu, Y.C. Lai, F.C. Hoppensteadt, Phys. Rev. E 63 
(2001)  055201.

\bibitem{Zheng1} Z. Zheng, B. Hu, G. Hu, Phys. Rev. E 62  (2000)  402-.

\bibitem{Hu} G. Hu, Y. Zhang, H.A. Cerdeira, S. Chen, Phys. Rev. Lett.
85  (2000) 3377-.

\bibitem{Zheng2} Z. Zheng, B. Hu, G. Hu, Phys. Rev. E 62  (2000)  402-.

\bibitem{manna} S.S. Manna, 
Phys. Rev. E 68 (2003) 027104.

\bibitem{garciamolina} M. Bawa, H. Garcia-Molina, A. Gionis, R. Motwani, 
Technical report, Stanford, 2003, submitted for publication; $%
http://dbpubs.stanford.edu/pub/2003-24$ 

\bibitem{Knyazeva2} H. Knyazeva, 
J.  Gen.  Philoso. Sci. 36 (2005) 289-304. 
\end{thebibliography}
\end{document}